\documentclass{PoS}

\title{Particle production vs. energy: how do simulation results match
experimental measurements?}

\ShortTitle{Particle production vs. energy.} 

\author{\speaker{M.BONESINI}%
         \\
        Sezione INFN Milano Bicocca\\
        E-mail: \email{maurizio.bonesini@mib.infn.it}}

\abstract{This talk is about the available hadroproduction
data to be used for the tuning of 
neutrino beamline simulations, going from conventional neutrino beams
to superbeams and neutrino factories.
Comparisons of data with existing Monte Carlo computations will be shown. 
Recent results from the HARP experiment at CERN PS will be summarized,
including data useful for a better understanding of extended air showers
 and atmospheric
neutrinos simulations.}

\FullConference{10th International Workshop on Neutrino Factories, Super beams and Beta beams\\
		 June 30 - July 5 2008\\
		 Valencia, Spain}

\begin{document}
\section{Introduction}
 The simulation of a conventional $\nu_{\mu}$ beam with a Monte Carlo (MC)
 is a delicate task due to complicate cascade
 processes involved in the neutrino production. 
 The paucity of
 available hadroproduction data, needed for MC tuning,  can limit systematically the precision 
 in the calculations.
 In addition, new hadroproduction data at low energies 
 are of great interest for extended air showers (EAS) and
 atmospheric neutrinos simulations
 and for the neutrino factory (NF) design. 
 One relevant point is how  existing MC simulations compare to available hadroproduction data.
 The more recent  dedicated hadroproduction experiments in the field are NA56/SPY~\cite{ref:spy}, 
MIPP~\cite{ref:MIPP}, NA61~\cite{ref:na61}  and HARP~\cite{harp}.
At low energies ($\leq 15$ GeV), the main experimental results come from the HARP experiment
at CERN PS and will be briefly summarized here. At higher energies we refer to \cite{ref:physrep} 
for further details. 
\section{The HARP experiment at CERN PS}
\begin{figure}[htb]
\vskip -0.8cm
\begin{center}
\includegraphics[width=0.5\linewidth]{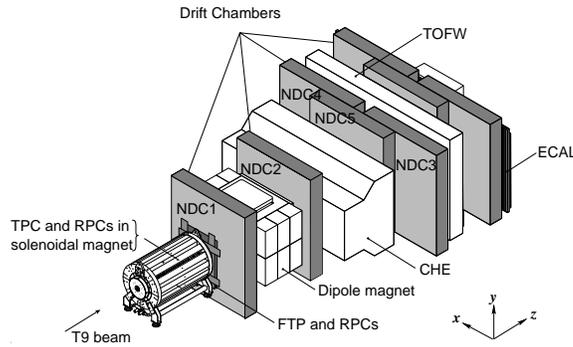}
\end{center}
\vskip -0.5cm
\caption{ Layout of the HARP experiment at CERN PS.}
\label{fig:harp}
\end{figure}

The HARP experiment at CERN PS \cite{harp} was designed to study  
hadroproduction
on nuclear targets (from $H_2$ to Ta) in the incident momentum range between 3 and 15 GeV/c.
The HARP detector is shown in Fig.~\ref{fig:harp} and includes different
subdetectors for tracking and particle identification (PID) over the full
solid angle. At large angle ($20^0 \leq \theta \leq 160^0$) tracking and PID
are performed by a TPC and an array of RPC counters. In the forward direction ($\theta
\leq 14.3^0$ ) the tracking device is a set of drift chambers, recovered
from the previous NOMAD experiment, while the PID is provided by a threshold
Cerenkov counter, a time of flight wall (TOFW) and an e.m. calorimeter~\cite{harp-1}.
Beam particles are tagged by a system of beam TOF detectors (TOFA,TOFB) and
Cherenkov counters.
Data were taken in 2001 and 2002, for a total of about 420 M triggers 
in $\sim 300$ experimental settings. 

\section{Results for simulation of NF beams.}

\begin{figure}[htb]
\begin{center}
\includegraphics[width=0.35\linewidth]{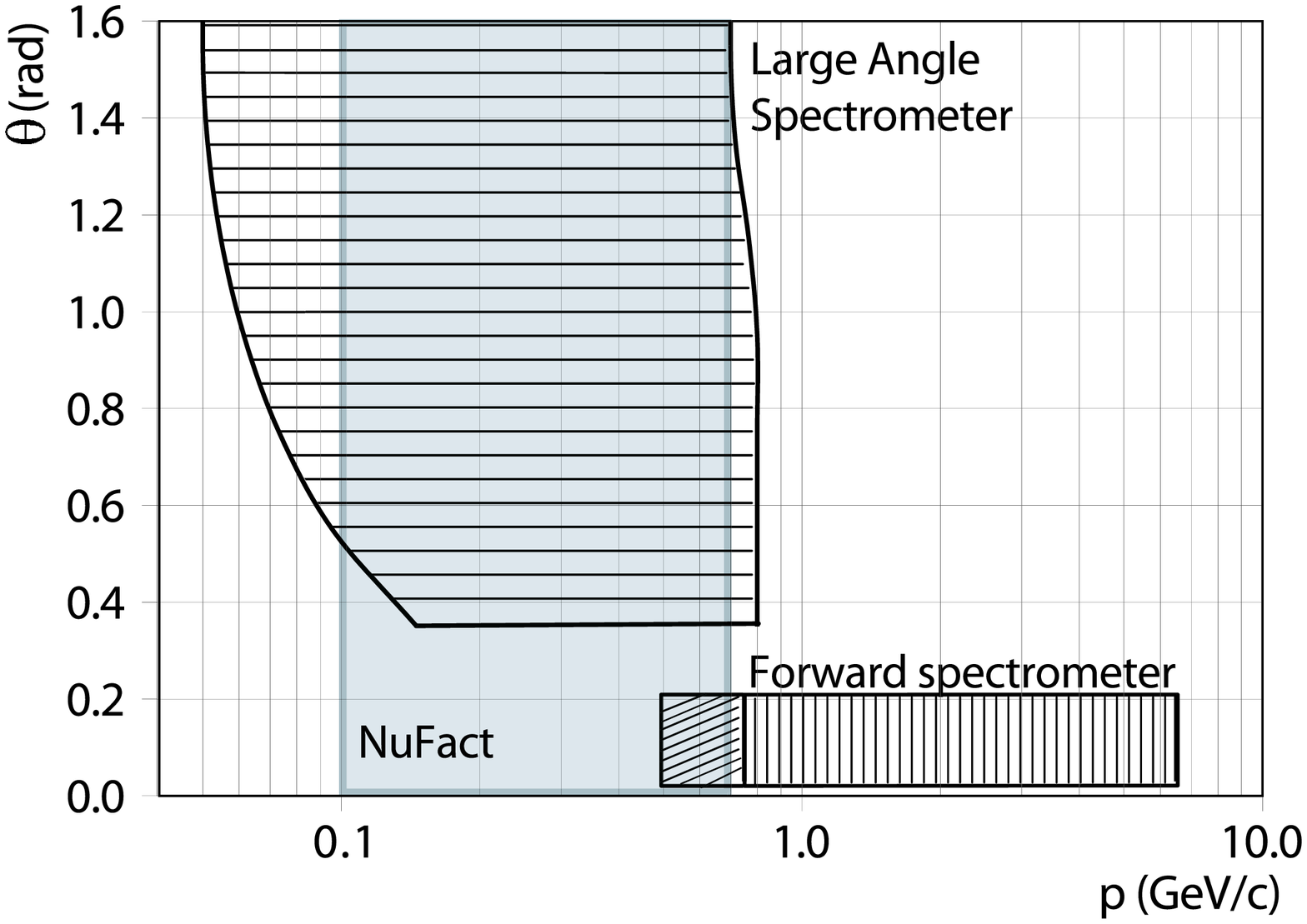}
\includegraphics[width=0.3\linewidth]{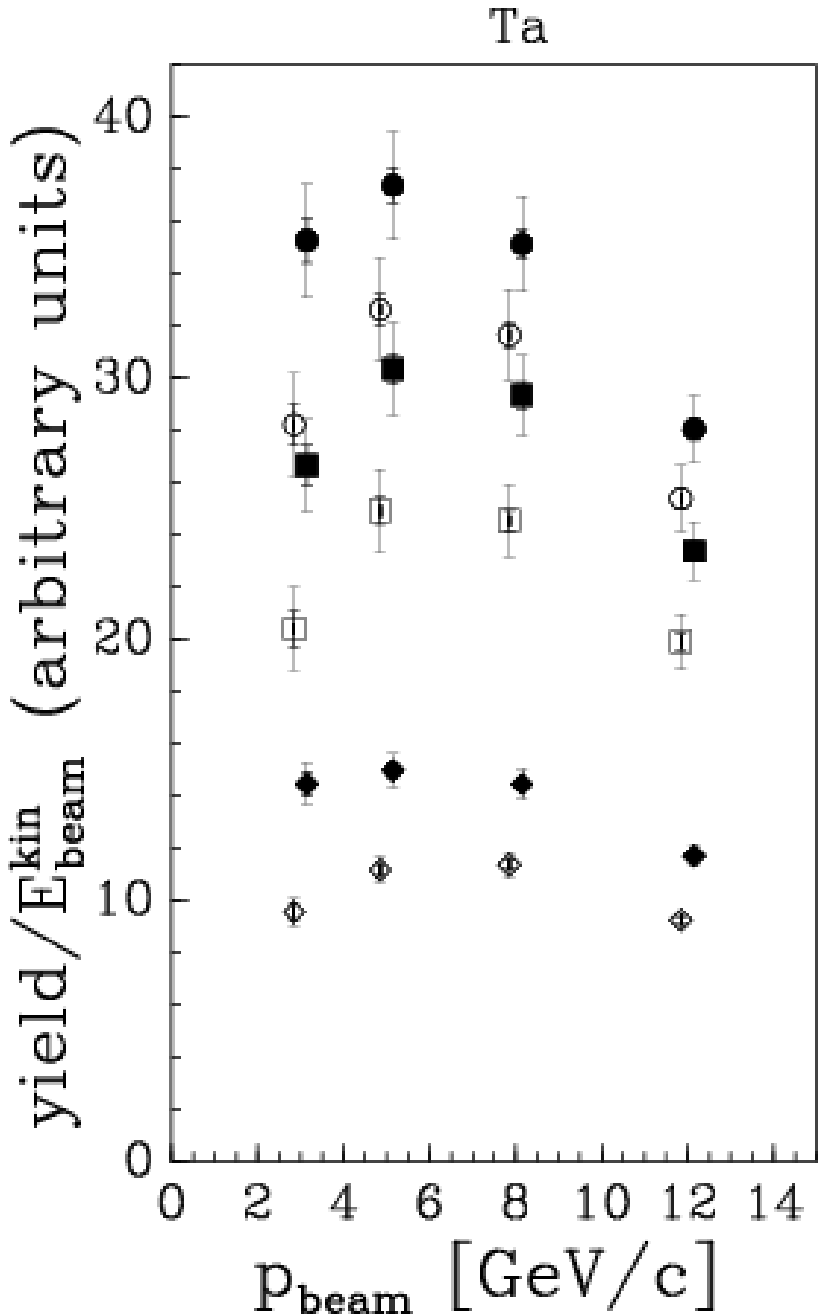}
\end{center}
\caption{Left panel: kinematic region in the $p-\theta$ plane covered by HARP
as compared to the acceptance of the input stage of typical NF designs.
Right panel:  $\pi^+$ (closed symbols) and $\pi^-$ 
(open symbols) yields as a function of the incident proton beam momentum
for different design of the NF focussing stage. The circles indicate the
integral over the full HARP acceptance, the squares are integrated
over 0.35 rad $\leq \theta \leq \ $ 0.95 rad, while the diamonds require in addition
the momentum cut 250 MeV/c $\leq p \leq $ 500 MeV/c.}
\label{fig:NF}
\end{figure}

\begin{figure}[htb]
\begin{center}
\includegraphics[width=0.45\linewidth]{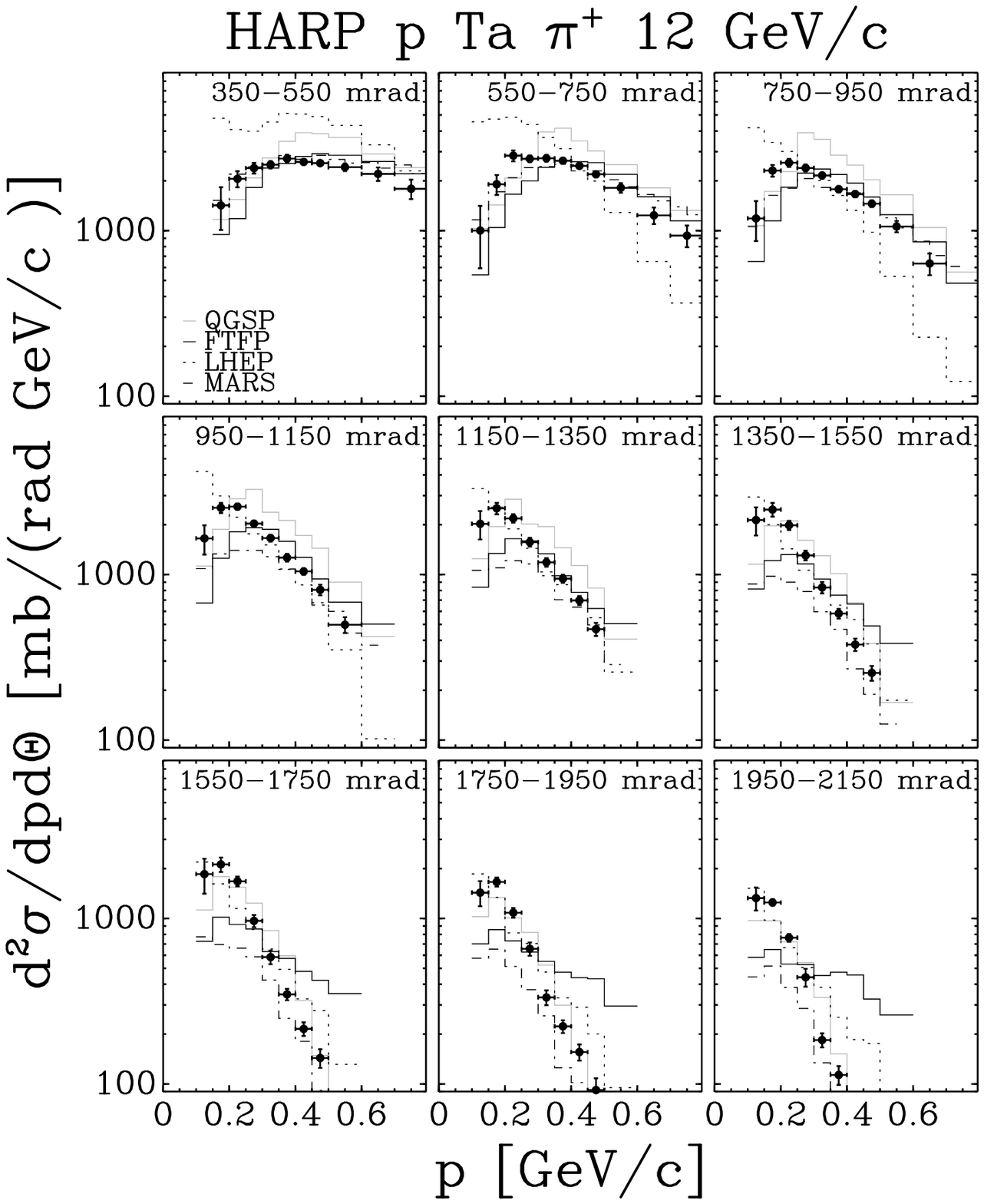}
\includegraphics[width=0.45\linewidth]{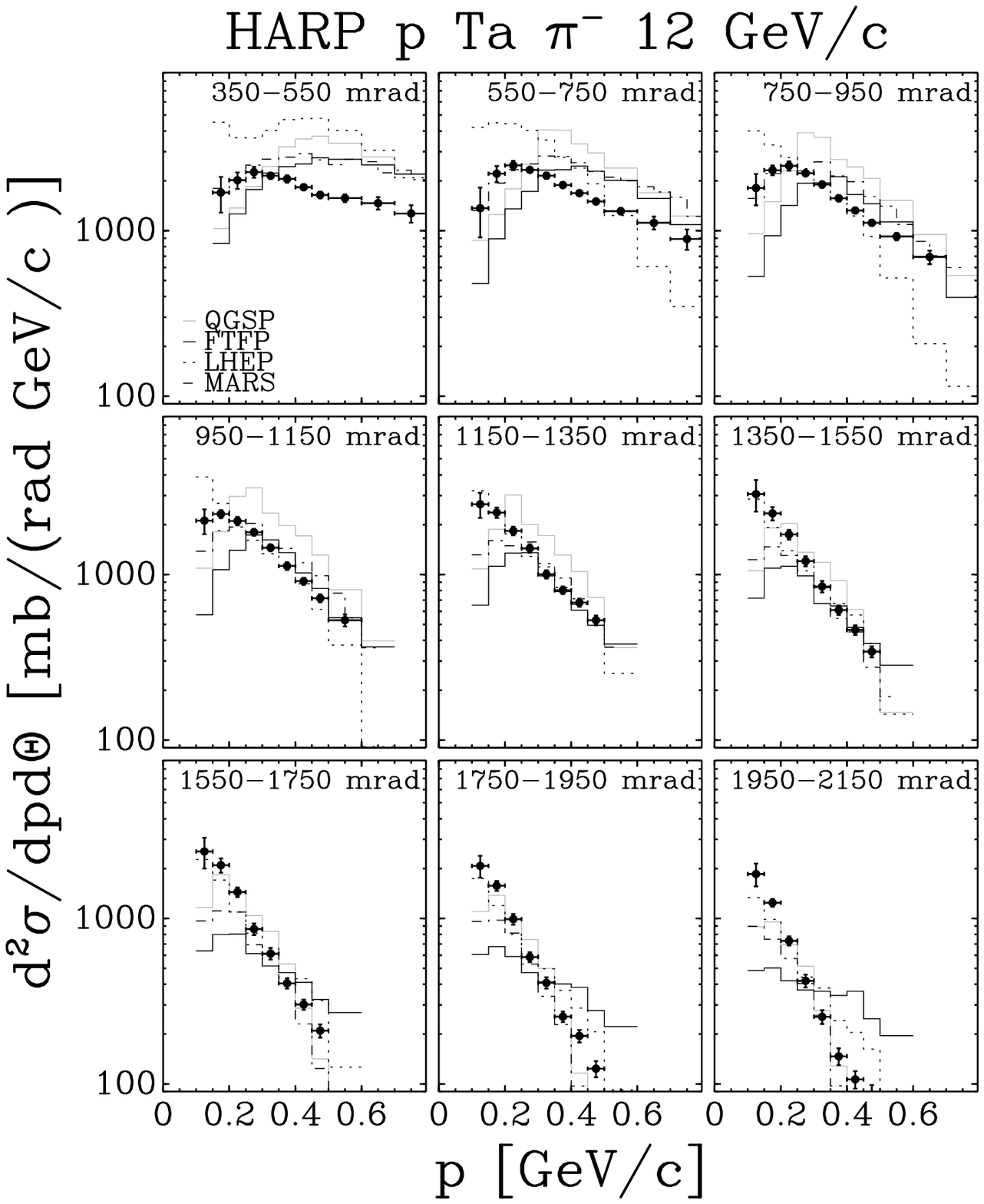}
\end{center}
\caption{Experimental results from HARP at 12 GeV/c for p-Ta cross sections
for $\pi^{\pm}$ production, as compared to MC models. See \cite{harp:LA} 
for further details and more data-MC comparisons.}
\label{fig:Ta}
\end{figure}

The baseline option for a NF target is a Hg jet target with impinging
particles at energies $10 \pm 5 $ GeV. Available data are very scarce 
and for the MC tuning the HARP data on heavy targets, such as Ta or Pb,
are of utmost importance.
The kinematical coverage of the HARP experiment is compared with the
acceptance of a typical NF design in figure \ref{fig:NF}. The experiment
covers the full momentum range of interest for production angles bigger than
0.35 rad.
The pion yield increases linearly with momentum and has
an optimum between 5 GeV/c and 8 GeV/c, as can be seen in 
the right panel of figure \ref{fig:Ta}.
Final results for pion production on heavy targets 
have been published in reference~\cite{harp:LA} and some comparisons
with available MC simulations are outlined in figure \ref{fig:Ta}. 
None of the considered models describe fully HARP data. However,
$\pi^+$ production is described better than $\pi^-$ production.
At lower (higher) energies binary and Bertini models from 
GEANT4 (the FTP model from GEANT4 and MARS) seem more appropriate.
Parametrized models (such as LHEP from GEANT4)  show relevant
discrepancies, up to a factor 3.

\section{Results for simulation of EAS and atmospheric neutrinos}
\begin{figure}[htb]
\begin{center}
\includegraphics[width=0.30\linewidth]{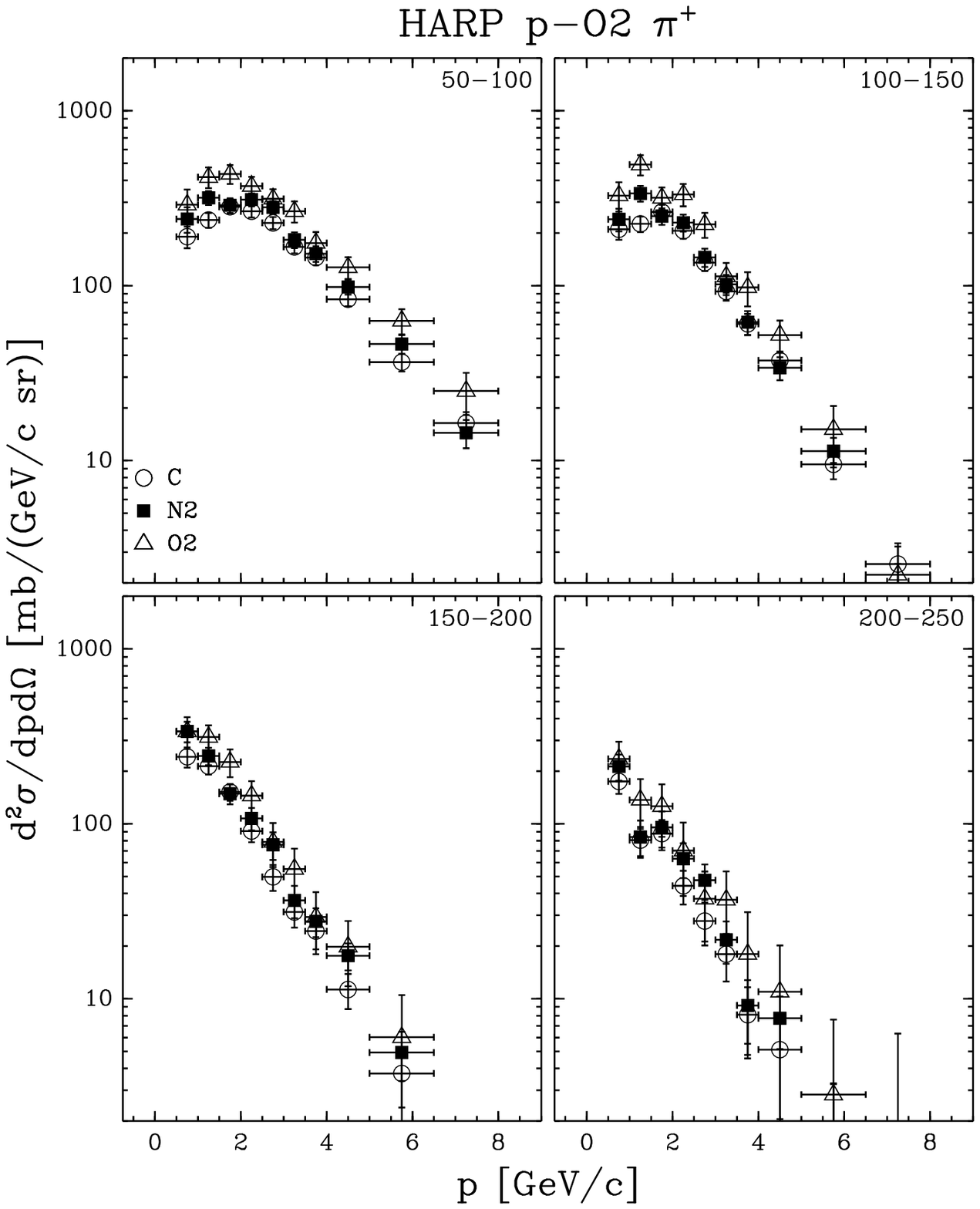}
\includegraphics[width=0.30\linewidth]{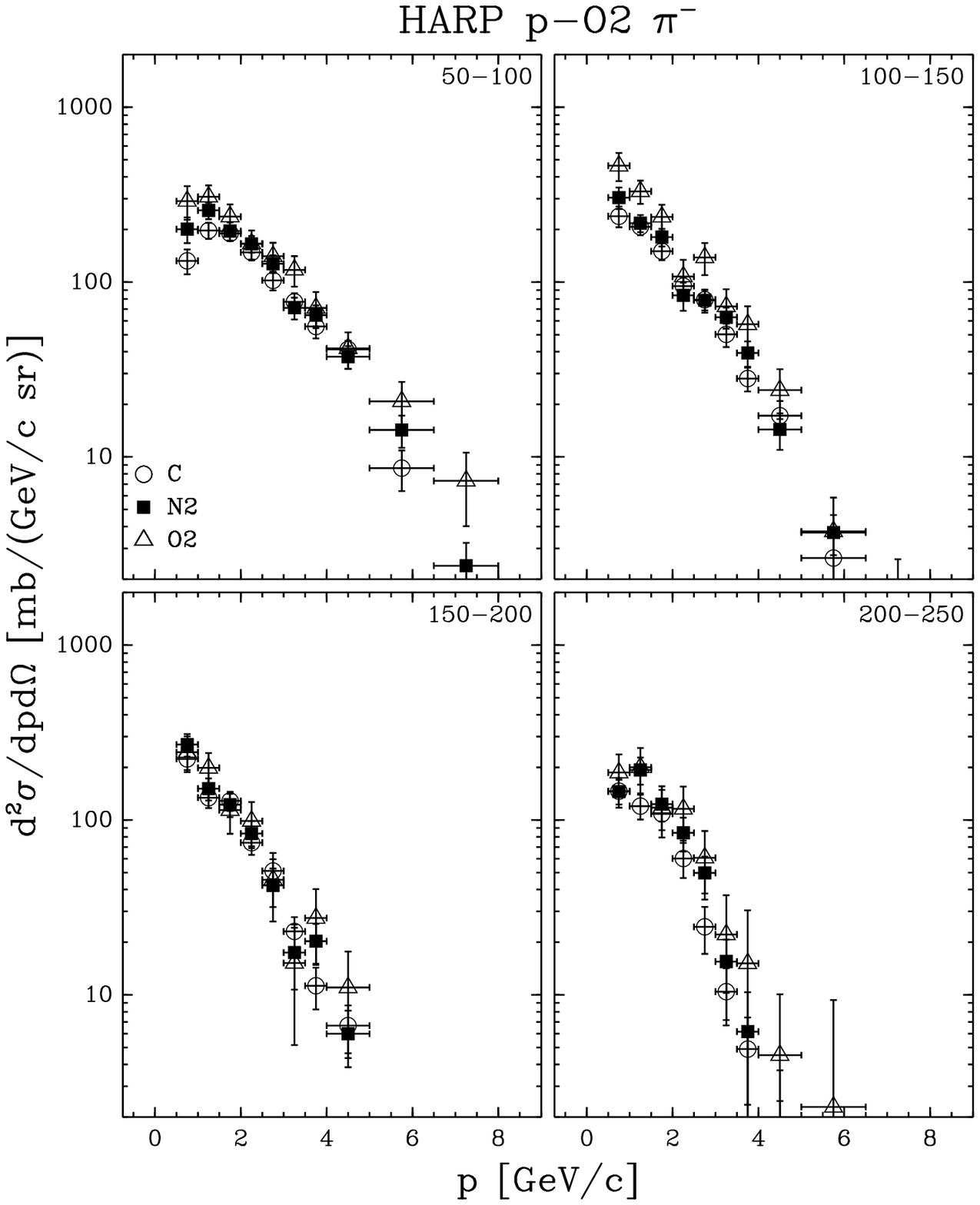}
\includegraphics[width=0.38\linewidth]{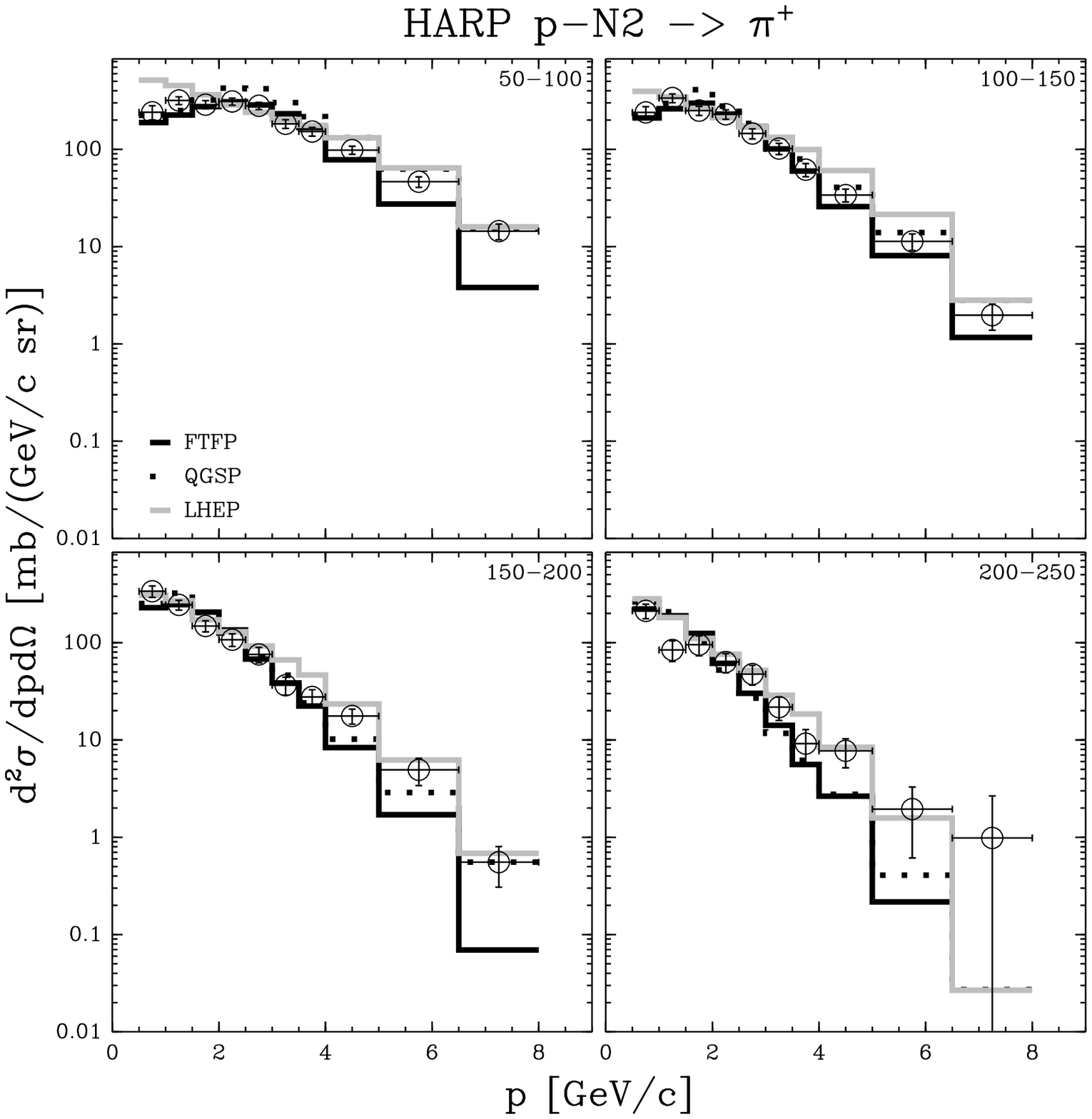}
\end{center}
\caption{Left: p--$O_2$,p--$N_2$,p--C cross sections at 12 GeV/c; 
right: comparison of $\pi^+$ production  in p--$N_2$ interactions with 
different MC models from GEANT4.}
\label{fig:N2}
\end{figure}

Results on cryogenic targets, such as $N_2$ and $O_2$ have a
direct impact on the precise
calculation of atmospheric neutrino fluxes and on the improved
reliability of extensive air shower simulations by reducing
the uncertainties of hadronic interaction models in the low energy range.
In particular, the common hypothesis that p--C
 data can be used to predict the p--N$_2$ and p--O$_2$ pion production
cross-sections may be tested.
HARP has published results \cite{ref:CNO} on charged pion production
cross-sections in interactions of 12~GeV/c protons on C, O$_2$ and N$_2$
thin targets, in the kinematic range
0.5~GeV/c  $\leq p_{\pi} <$ 8~GeV/c
and 50~mrad $\leq \theta_{\pi} <$ 250~mrad. Some results, showing also a 
comparison with available simulations, are reported in figure~\ref{fig:N2}.

\section{Results for simulation of conventional neutrino beams:
the MiniBooNE and K2K physics cases.}

Prediction of the far detector spectrum in the absence of oscillations
is a key ingredient in a neutrino oscillation experiment. This
can be done by an extrapolation from a near detector via a nominal
far/near ratio estimated by a beamline MC simulation.
The error on the observed number of events in the K2K far detector
(SuperKamiokande) was dominated by contributions from uncertainties of normalization
($\pm 5 \%$) and far/near ratio ($\pm 5 \%$).

HARP has reported measurements of the $\pi^+$ production in p--Al interactions 
at 12.9 GeV/c \cite{ref:Al}. These results have 
contributed in a significant way to reduce 
the systematic error associated to the FAR/NEAR ratio, thus increasing the
K2K sensitivity to oscillation signals \cite{ref:K2K}.
 
\begin{figure}[htb]
\begin{center}
\includegraphics[width=0.6\linewidth]{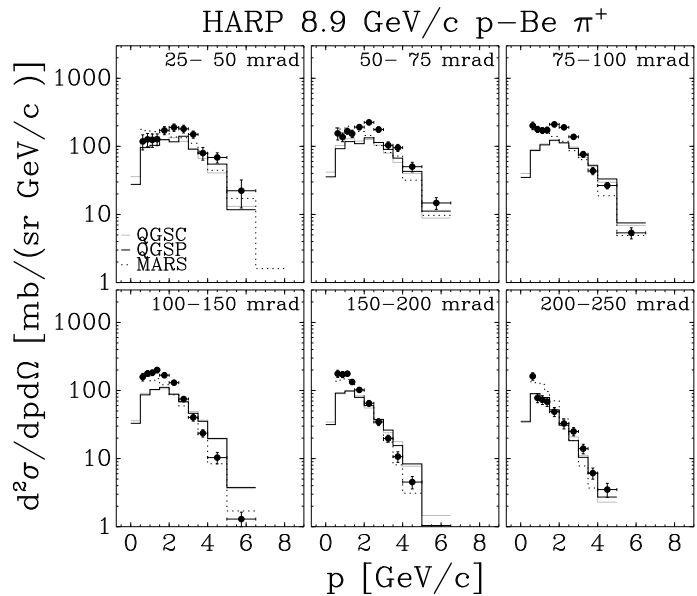}
\end{center}
\caption{Experimental results from HARP at 8.9 GeV/c for p-Be cross sections
for $\pi^{+}$ production, as compared to MC models from GEANT4 (QGSC,QGSP)
and MARS.}
\label{fig:Be}
\end{figure}

Similar results were obtained in 8.9 GeV/c p--Be interactions \cite{ref:Be}
and have contributed to a better understanding of the MiniBooNE and SciBooNE
$\nu$ fluxes.
Figure \ref{fig:Be} reports the comparison with some available MC models.

%


\begin{thebibliography}{9}
\bibitem{ref:spy} G. Ambrosini {\it et al.}, NA56/SPY Collaboration, Eur. J. Phys. C10 (1999) 605 ; \\
M. Bonesini {\it et al.} Eur. J. Phys. C20 (2001) 13. 
\bibitem{ref:MIPP} MIPP Collaboration, http://ppd.fnal.gov/experiments/2907.
\bibitem{ref:na61} N. Antoniou {\it et al.}, NA61 Collaboration, CERN-SPSC-P-330, 2006.
\bibitem{harp} M.G. Catanesi {\it  et al.}, 
 CERN-SPSC/99-35, SPSC/P315, 15 November 1999.
\bibitem{ref:physrep} M. Bonesini, A. Guglielmi, Phys. Rep. 433 (2006) 65.
\bibitem{harp-1} M. G. Catanesi {\it  et al.},
Nucl. Instr. Meth. A571 (2007) 527; \\
 M. Baldo-Ceolin  {\it et al.}, 
Nucl. Instr. Meth.  A532 (2004) 548.
\bibitem{harp:LA} M.G. Catanesi {\it et al.}, HARP Collaboration, Phys. ReV. C77 (2008) 0555207 \\
M.G. Catanesi {\it et al.}, HARP Collaboration, Eur. Phys. J. C51 (2007) 787.
\bibitem{ref:CNO} M.G. Catanesi {\it et al.}, HARP Collaboration, Astr. Phys. 29 (2008) 257; \\
M.G. Catanesi {\it et al.}, HARP Collaboration, `` Forward $\pi^{\pm}$ production in p--$O_2$ and 
p--$N_2$ interactions at 12 GeV/c'', in press on Astr. Phys.
\bibitem{ref:Al} M.G. Catanesi {\it et al.}, HARP Collaboration, Nucl. Phys. B732 (2006) 1.
\bibitem{ref:K2K} M.H. Ahn {\it et al.}, K2K Collaboration, Phys. ReV.D74 (2006) 072003.
\bibitem{ref:Be} M.G. Catanesi {\it et al.}, HARP Collaboration, 
Eur. Phys. J. C52 (2007) 49.



\end{thebibliography}
\end{document}